\documentclass[runningheads]{llncs}

\usepackage{svg}
\usepackage{mathrsfs}
\usepackage{hyperref}

\usepackage{graphicx}
\usepackage{tabularx}
\usepackage{bibnames}
\usepackage{subcaption}
\usepackage{mhchem}
\begin{document}
\title{An Ensemble Approach for Patient Prognosis of Head and Neck Tumor Using Multimodal Data}
%
%
\author{Numan Saeed\and
Roba Al Majzoub\and
Ikboljon Sobirov\and
Mohammad Yaqub
}
\authorrunning{N. Saeed et al.}
%
\titlerunning{Multimodal Prognosis Model for H\&N Tumor}
\institute{Mohamed bin Zayed University of Artificial Intelligence, Abu Dhabi, UAE
\\
\url{https://mbzuai.ac.ae/biomedia} \\
\email{\{numan.saeed, roba.majzoub, ikboljon.sobirov, mohammad.yaqub\}@mbzuai.ac.ae}}
\maketitle              
\begin{abstract}

Accurate prognosis of a tumor can help doctors provide a proper course of treatment and, therefore, save the lives of many. Traditional machine learning algorithms have been eminently useful in crafting prognostic models in the last few decades. Recently, deep learning algorithms have shown significant improvement when developing diagnosis and prognosis solutions to different healthcare problems. However, most of these solutions rely solely on either imaging or clinical data. Utilizing patient tabular data such as demographics and patient medical history alongside imaging data in a multimodal approach to solve a prognosis task has started to gain more interest recently and has the potential to create more accurate solutions. The main issue when using clinical and imaging data to train a deep learning model is to decide on how to combine the information from these sources. We propose a multimodal network that ensembles deep multi-task logistic regression (MTLR), Cox proportional hazard (CoxPH) and CNN models to predict prognostic outcomes for patients with head and neck tumors using patients' clinical and imaging (CT and PET) data. Features from CT and PET scans are fused and then combined with patients' electronic health records for the prediction. The proposed model is trained and tested on 224 and 101 patient records respectively. Experimental results show that our proposed ensemble solution achieves a C-index of 0.72 on The HECKTOR test set that saved us the first place in prognosis task of the HECKTOR challenge. The full implementation based on PyTorch is available on \url{https://github.com/numanai/BioMedIA-Hecktor2021}.       

\textbf{Team name:} MBZUAI-BioMedIA

\end{abstract}
\section{Introduction}
Each year, 1.3 million people are diagnosed with head and neck (H\&N) cancer worldwide on average \cite{wang}. However, the mortality rate can be lowered to 70\% with early detection of H\&N tumor \cite{wang}. Therefore, diagnosis and prognosis are the two primary practices involved in most medical treatment pipelines, especially for cancer-related diseases. After determining the presence of cancer, a doctor tries to prescribe the best course of treatment yet with limited information, it is very challenging. An early survival prediction can help doctors pinpoint a specific and suitable treatment course. Different biomarkers from radiomics field can be used to predict and prognose medical cases in a non-invasive fashion \cite{robert}. It is used in oncology to help with cancer prognosis, allowing patients to plan their lives and actions in their upcoming days. In addition, it enables doctors to better plan for the time and mode of action followed for treatment \cite{mack}. This is necessary to make more accurate predictions, which, in turn, is likely to lead to better management by the doctors. 

Many other research fields also strive to assist medical doctors, at least to a point of alleviating their work process. One of the most common statistical frameworks used for the prediction of the survival function for a particular unit is the Cox proportional hazard model (CoxPH), proposed by Cox \cite{cox} in 1972. It focuses on developing a hazard function, i.e., an age-specific failure rate. Nevertheless, CoxPH comes with specific issues, such as the fact that the proportion of hazards for any two patients is constant or that the time for the function is unspecified. Yu et al. \cite{yu} proposed an alternative to CoxPH - multi-task logistic regression (MTLR). MTLR can be understood as a sequence of logistic regression models created at various timelines to evaluate the probability of the event happening. Fotso \cite{fotso} improved the MTLR model by integrating neural networks to achieve nonlinearity, yielding higher results.

Deep learning (DL) has gained a considerable amount of attention in classification, detection, and segmentation tasks of the medical research field. Furthermore, their use in far more complicated tasks such as prognosis and treatment made DL even more popular, as it can handle data in large amounts and from different modalities, both tabular and visual. 

Many studies have been conducted to perform prognosis of cancer using DL. Sun et al. \cite{sun} propose a deep learning approach for the segmentation of brain tumor and prognosis of survival using multimodal MRI images. 4524 radiomic features are extracted from the segmentation outcome, and further feature extraction is performed with a decision tree and cross-validation. For survival prediction, they use a random forest model. In a similar task done by Shboul et al. \cite{shboul}, a framework for glioblastoma and abnormal tissue segmentation and survival prediction is suggested. The segmentation results, along with other medical data, are combined to predict the survival rate. Tseng et al. \cite{tseng} develop a multiclass deep learning model to analyze the historical data of oral cancer cases. They achieve superior results compared to traditional logistic regression.

A few studies have been conducted for the prognosis of H\&N cancer \cite{kazmierski}. The prognosis studied in \cite{parmar} shows that they achieve the area under the curve (AUC) of 0.69 for their best-performing dataset for H\&N tumor, while for the rest of the datasets, they achieve AUC between 0.61 and 0.68. Furthermore, Kazmierski et al \cite{kazmierski} use electronic health record (EHR) data and pre-treatment radiological images to develop a model for survival prediction in H\&N cancer. Out of the many trials they experimented with, a non-linear, multitask approach that uses the EHR data and tumor volume produced the highest result for prognosis. 

Such results to this task are unlikely to motivate clinicians to use machine learning models in clinical practice; therefore, more accurate prognosis is critical to help solve this problem .  In this paper, we are proposing a multimodal machine learning algorithm that, without prior information on the exact location of the tumor, utilizes both tabular and imaging data for the prognosis of Progression Free Survival (PFS) for patients who have H\&N oropharyngeal cancer. This work is carried out to address the prognosis task of the MICCAI 2021 Head and Neck Tumor segmentation and outcome prediction challenge (HECKTOR) \cite{hecktor_new}\cite{hecktor_new2}.

\section{Materials and Methods}

\subsection{Dataset}
\begin{figure}[h!]
\captionsetup[subfigure]{justification=centering}
\centering
\begin{minipage}{0.3\textwidth}
\begin{subfigure}{\textwidth}
    \includegraphics[height=3.5cm,width=\textwidth]{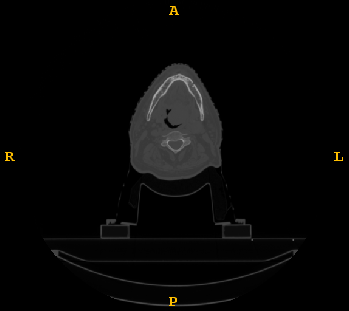}
    \subcaption{\textbf{CT}}
\end{subfigure}
\end{minipage}
\begin{minipage}{0.3\textwidth}
\begin{subfigure}{\textwidth}
    \includegraphics[height=3.5cm,width=\textwidth]{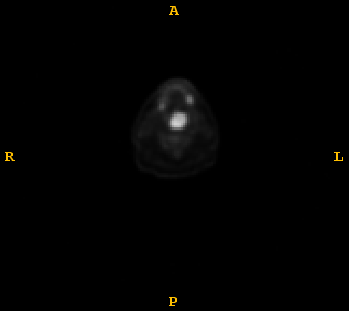}
    \subcaption{\textbf{PET}}
\end{subfigure}
\end{minipage}
\begin{minipage}{0.3\textwidth}
\begin{subfigure}{\textwidth}
    \includegraphics[height=3.5cm,width=\textwidth]{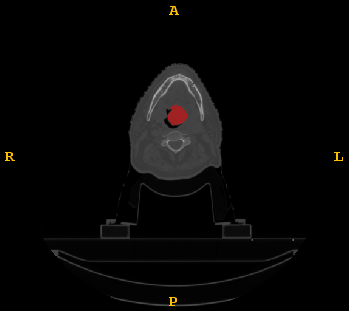}
    \subcaption{\textbf{Mask}}
\end{subfigure}
\end{minipage}
\caption{Examples from the training set of HECKTOR 2021 \cite{hecktor_new2}. (a) shows the CT slice, (b) shows the PET slice, and (c) shows the mask (in red) superimposed on the CT image.}
\label{fig:samples}
\end{figure}

\noindent The HECKTOR committee \cite{hecktor_new}\cite{hecktor_new2} provided CT and PET scans, manual segmentation masks and electronic health record (EHR) dataset. The ground truth segmentation masks for H\&N oropharyngeal cancer was manually delineated by oncologists as shown in Fig. \ref{fig:samples}. EHR contains data about a patient's age, gender, weight, tumor stage, tobacco and alcohol consumption, chemotherapy experience, presence of human papillomavirus (HPV) and other data. A clinically relevant endpoint was provided in the training set to predict PFS for each patient. The data is multi-centric, collected from four centers in Canada, one center in Switzerland and another one in France. The total number of patients involved in the study was 325, out of which 224 were training cases and 101 test cases. However, some data points are missing in the EHR, e.g., tobacco and alcohol data were only provided by one of the centers mentioned above in the training set. The dataset has 75\% of the patients censored, who might have stopped following up or changed the clinic. 

\begin{figure}[h!]
\captionsetup[subfigure]{justification=centering}
\centering
\begin{minipage}{0.32\textwidth}
\begin{subfigure}{\textwidth}
    \includegraphics[height=2.8cm,width=\textwidth]{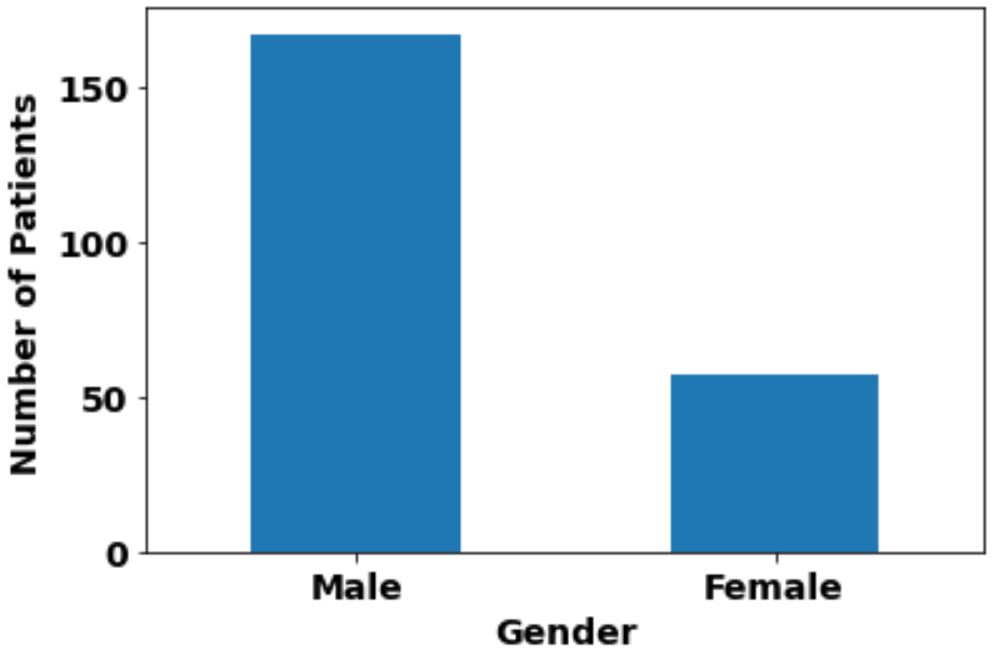}
   \end{subfigure}
\end{minipage}
\begin{minipage}{0.32\textwidth}
\begin{subfigure}{\textwidth}
    \includegraphics[height=2.8cm,width=\textwidth]{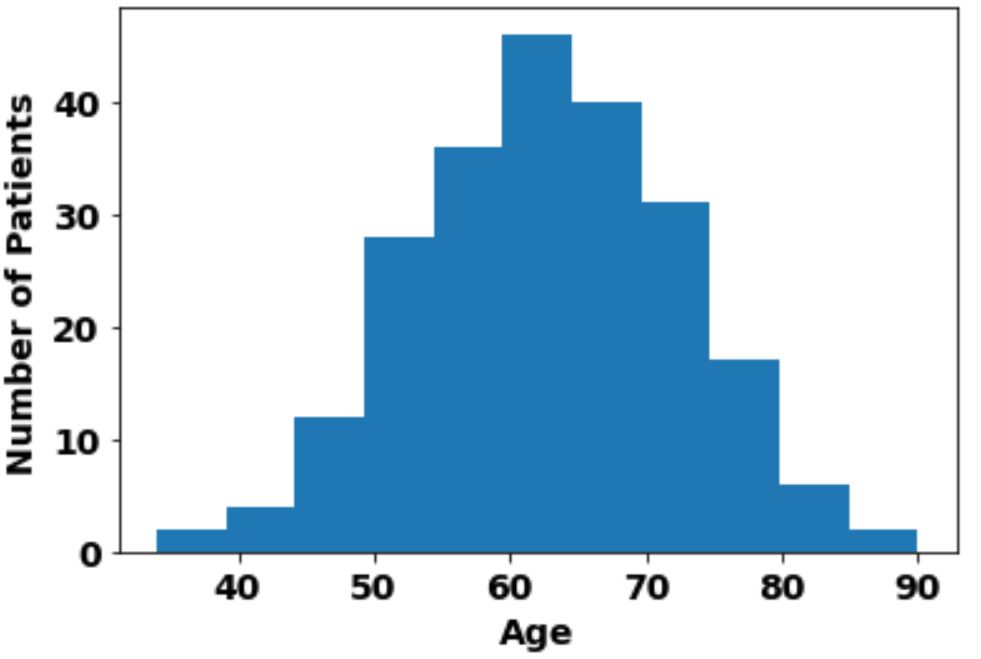}
    \end{subfigure}
\end{minipage}
\begin{minipage}{0.32\textwidth}
\begin{subfigure}{\textwidth}
    \includegraphics[height=2.8cm,width=\textwidth]{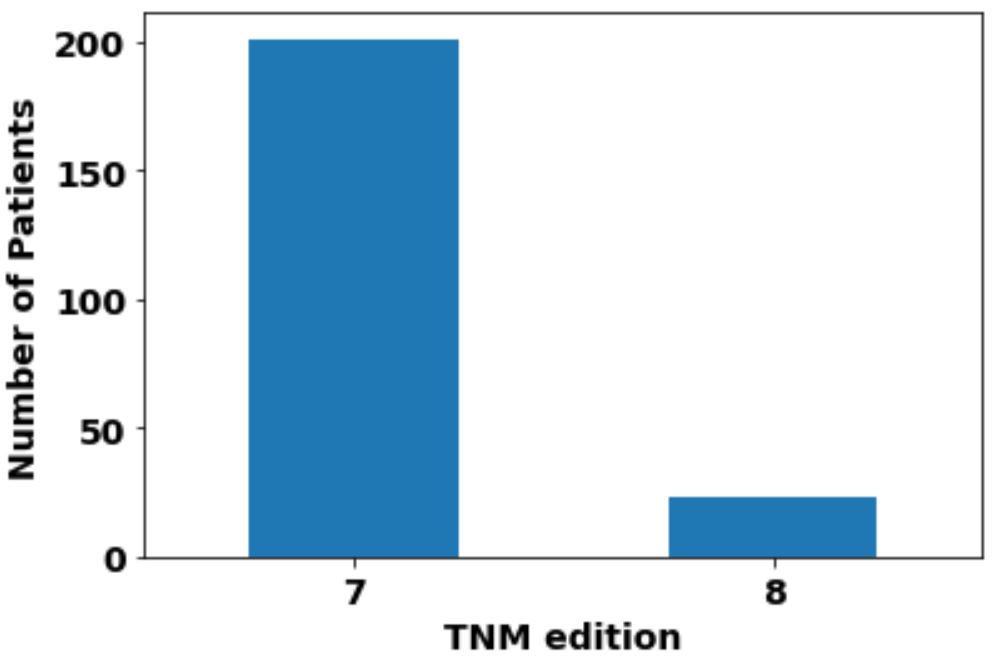}
    \end{subfigure}
\end{minipage}
\caption{EHR data visualization. The graph on the left shows the male-to-female ratio. The chart in the middle depicts the distribution of age. The chart on the right shows the TNM edition for all patients.}
\label{fig:vis}
\end{figure}

\noindent Visualization of the EHR data was performed to observe the distribution of patients in terms of gender and age, as shown in Fig. \ref{fig:vis}. Most of the patients are males, and the age ranges between 35 to 90 years with the peak at around 60 years. The TNM (T: tumor, N: nodes, and M: metastases) edition 7 is used for most patients to describe the volume and spread of cancer in a patient's body, while the rest of the patients' cases were represented using TNM edition 8.

\subsection{Data Analysis and Image Preprocessing}
\begin{figure}[h!]
\begin{minipage}{0.48\textwidth}
\begin{subfigure}{\textwidth}
    \includegraphics[height=4cm,width=\textwidth]{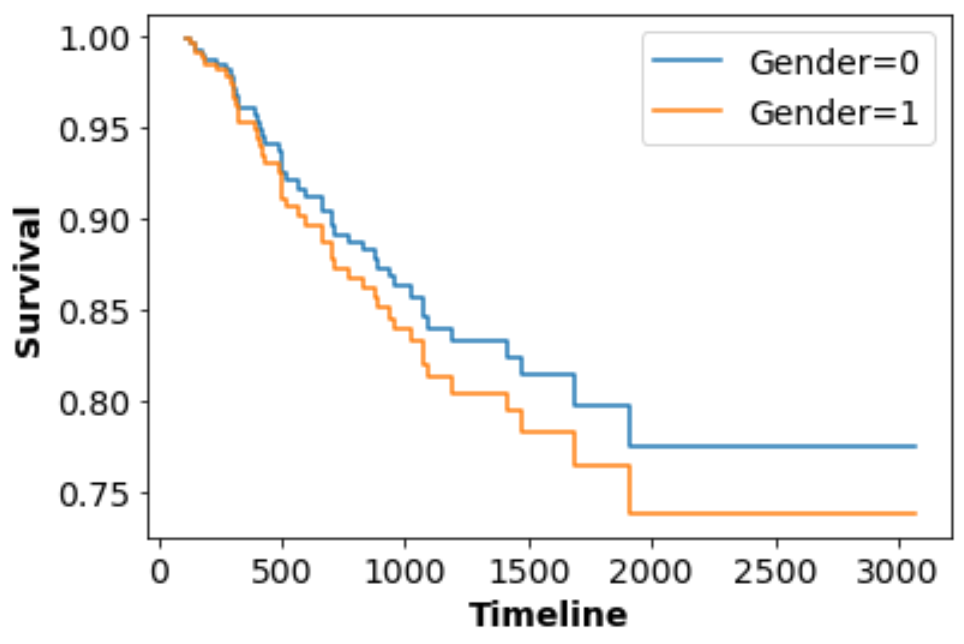}
    \subcaption{\textbf{Gender}}
\end{subfigure}
\end{minipage}
\begin{minipage}{0.48\textwidth}
\begin{subfigure}{\textwidth}
    \includegraphics[height=4cm,width=\textwidth]{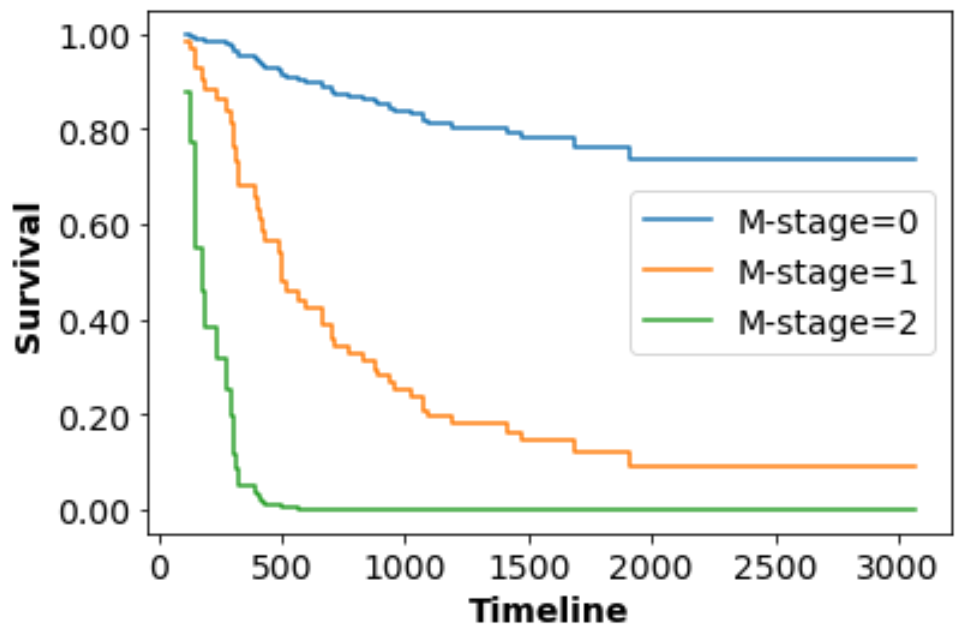}
    \subcaption{\textbf{Metastasis (M)}}
\end{subfigure}
\end{minipage}
\caption{Partial effects on outcome when covariate(s) is varied. (a) shows the survival rate for males and females. The baseline we compared to and males' survival prediction curves are superimposed in the figure. (b) depicts the metastasis effects on the survival of patients.}
\label{fig:CoxPH}
\end{figure}

  We initially analyzed the EHR data from the training dataset using the CoxPH model by splitting them into training, validation and testing sets to experiment on different hyperparameters and configurations of the solution. Then, we tried to observe the effects of different covariates on the survival rate using the trained CoxPH model. In Fig. \ref{fig:CoxPH} (a), the gender covariate is varied by assigning males and females the values 0 and 1 respectively. The results show that the survival rate of males is less than that of females. Similarly, to observe the effect of metastasis (M), M1, M2, and Mx are assigned the values 0, 1, and 2 respectively. The analysis shows that the patients with cancer spread to other parts of their bodies (M1) have lower survival rates, which is in line with the medical science. Some data points for tobacco and alcohol were not available, so we tried to impute them through assigning values of 1 for consumers, -1 for non-consumers and 0 for patients with missing tobacco and alcohol consumption data. Then, we tried another approach where we dropped the use of incomplete data features and trained the model on other available data features. The obtained results of the cross validation revealed that dropping incomplete data points leads to better prognosis results than imputing them.

\begin{figure}[h!]
\captionsetup[subfigure]{justification=centering}
\centering
\begin{minipage}{0.3\textwidth}
\begin{subfigure}{\textwidth}
    \includegraphics[height=3.5cm,width=\textwidth]{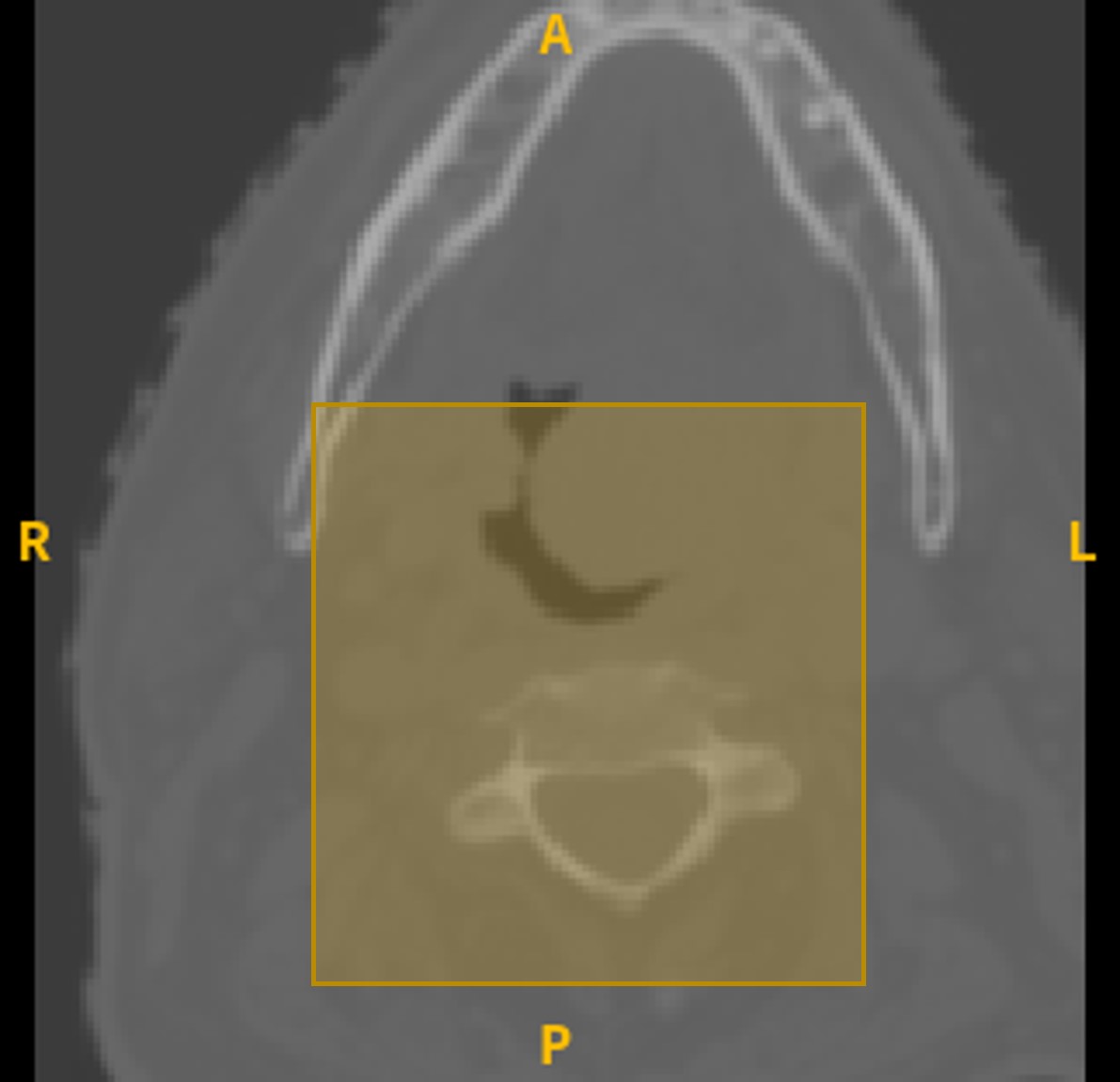}
    \subcaption{\textbf{CT}}
\end{subfigure}
\end{minipage}
\begin{minipage}{0.3\textwidth}
\begin{subfigure}{\textwidth}
    \includegraphics[height=3.5cm,width=\textwidth]{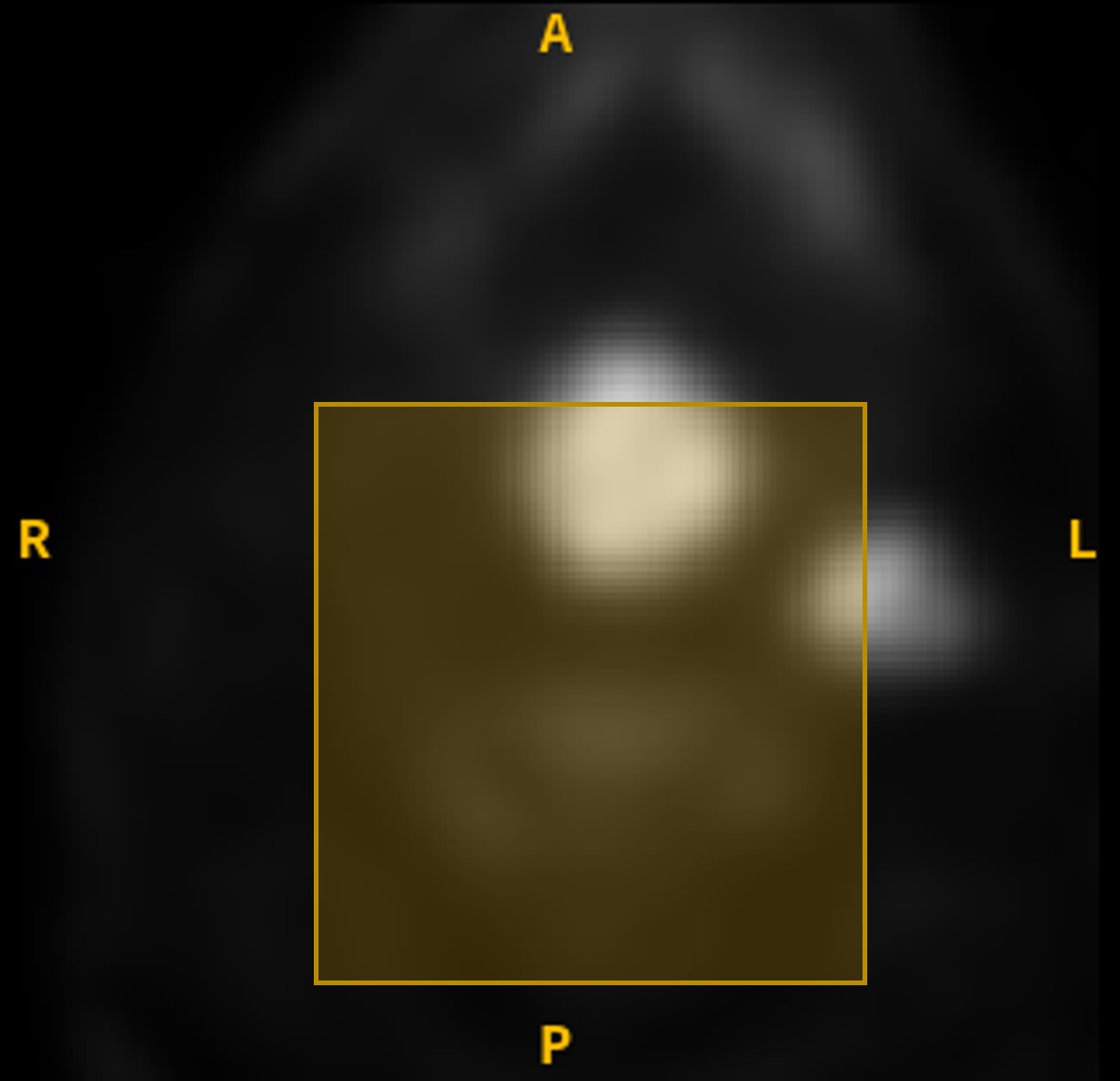}
    \subcaption{\textbf{PET}}
\end{subfigure}
\end{minipage}
\begin{minipage}{0.3\textwidth}
\begin{subfigure}{\textwidth}
    \includegraphics[height=3.5cm,width=\textwidth]{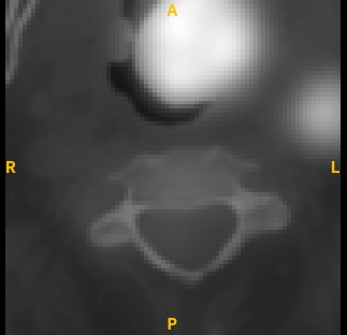}
    \subcaption{\textbf{Fused}}
\end{subfigure}
\end{minipage}
\caption{An example of combining CT and PET scans. (a) depicts the (144x144x144) CT scan with a rectangle to show the region to be cropped. (b) shows the (144x144x144) PET scan and the cropping region. (c) shows the (80x80x50) fused image in the cropped form.}
\label{fig:fusure}
\end{figure}

 As for the image dataset, PET and CT scans were preprocessed using the bounding box information available in the provided csv file to obtain 144x144x144 cropped images. To prepare the image data for our model input, we normalized the two images to the same scale and a fused image was created by averaging the two scans for each patient. To further reduce the volume, the fused output image is cropped again based on a specific distance away from the center of the 144x144x144 cube as shown in Fig. \ref{fig:fusure}. Two possible crop resolutions were considered: 50x50x50 and 80x80x50; however, the latter was adopted since it resulted in a better outcome.

\subsection{Network Architecture}
The baseline we proposed has an MTLR model as its cornerstone and was later expanded and developed by integrating and optimizing features obtained from different inputs. First, from the EHR data provided in the training dataset, a prognostic model was developed. Then, we optimized the vanilla MTLR by experimenting with different hyperparameters such as learning rates, the depth and width of the feedforward layers, and the constant $C$ in \textit{$l_2$} regularization term of the loss function of MTLR from \cite{jin2015using} depicted in the following equation:

\scriptsize{\[\min _{\Theta} \frac{C}{2} \sum_{j=1}^{m}\left\|\vec{\theta}_{j}\right\|^{2}-\sum_{i=1}^{n}\left[\sum_{j=1}^{m} y_{j}\left(s_{i}\right)\left(\vec{\theta}_{j} \cdot \vec{x}_{i}+b_{j}\right)-\log \sum_{k=0}^{m} \exp f_{\Theta}\left(\vec{x}_{i}, k\right)\right]\]}

\normalsize
The smoothness of the predicted survival curves depends on the change between consecutive timepoints and is controlled by $C$. 

Next, we investigated the effect of multimodality on the performance of the model by integrating the available image data. Features were extracted from the fused crops through the use of a 3D convolutional neural network (CNN) adopted from \cite{kim} named Deep-CR. Unlike \cite{kim}, we optimized the CNN architecture using \texttt{OPTUNA} framework \cite{optuna} to obtain the best hyperparameters, including the kernel sizes and the number of layers. These features were concatenated with our tabular data and fed into two fully connected layers. Finally, the risk was calculated using MTLR, and the output was averaged with CoxPH model risk output.

\begin{figure*}[h!]
    \centering
    \includegraphics[width=1\textwidth]{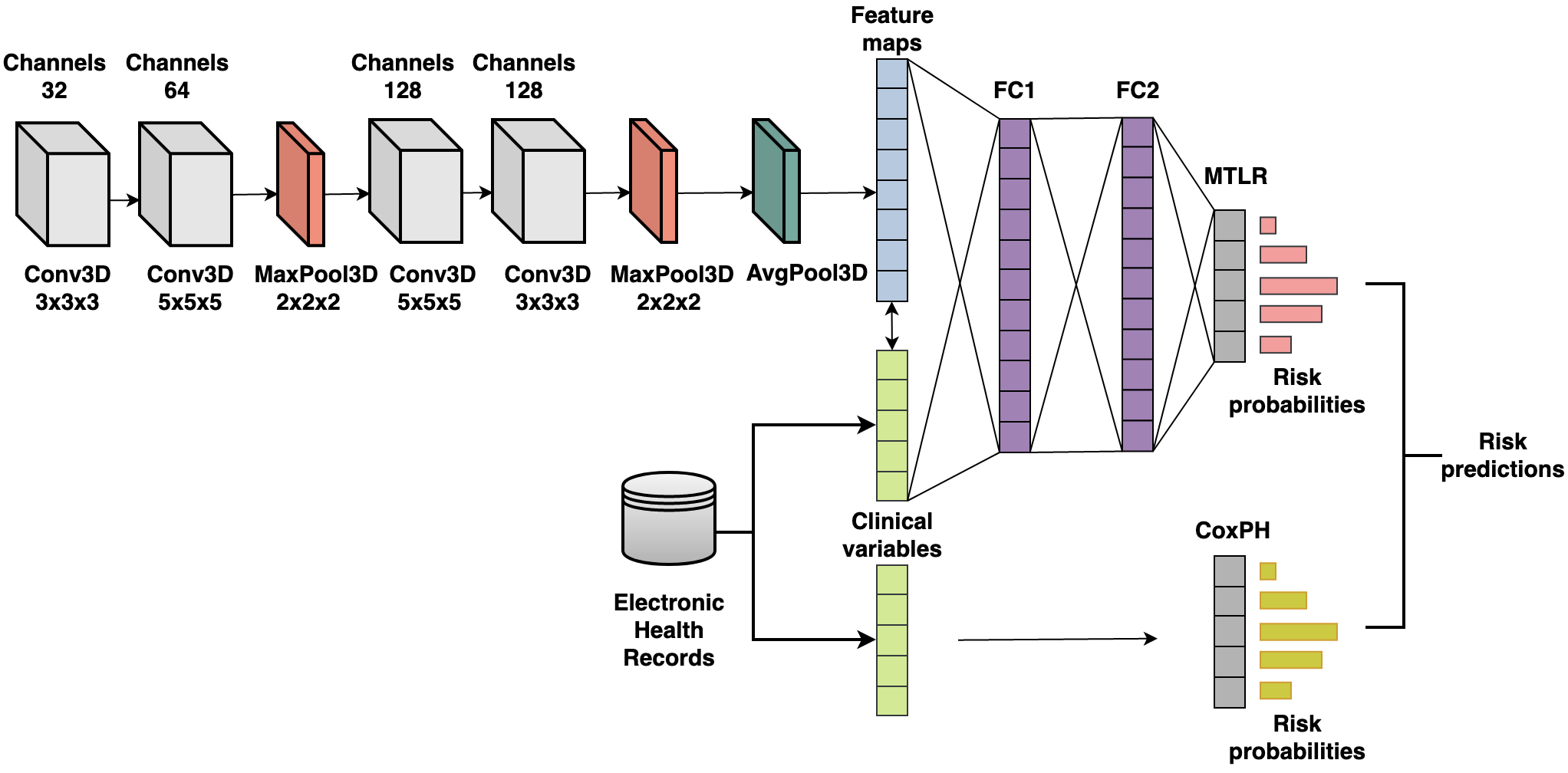}
    \caption{Overall Architecture of Deep Fusion V2. Features are extracted from fused CT and PET scans using the CNN network and concatenated with the EHR features. Then, the output is passed to the FC layers before MTLR. Lastly, risk scores from MTLR and CoxPH models are averaged to get the final risk predictions.}
    \label{fig:network}
\end{figure*}

For our best-performing model, inspired by the Deep-CR approach, we developed a network as shown in Fig. \ref{fig:network} to predict the individual risk scores. An optimized Deep-CR network with two blocks was developed. Each block consists of two 3D convolutional, ReLU activation and batch normalization layers. The 3D CNN blocks are followed by 3D MaxPooling layers. The kernel sizes of the 3D CNN layers in each block are 3 and 5. The number of output channels of the 3D CNN layers are 32, 64, 128 and 256 respectively as shown in the Fig. \ref{fig:network}. The number of neurons in the two feed forward layers are 256 each. The batch size, learning rate, and dropout were experimentally set to 16, 0.016, and 0.2 respectively for the training. The model was trained for 100 epochs using Adam optimizer on a single GPU NVIDIA RTX A6000 (48 GB).  We refer to this network as Deep Fusion, and we implement two variants of it; (V1) with three 3D CNN paths that take three types of image inputs (CT, PET, and fused images) and (V2), which includes one 3D CNN path with a single image input (the fused data) as shown in Fig. \ref{fig:network}. Each CNN outputs a feature vector of length 256. In Deep Fusion V1, the 3 feature vectors are concatenated. The feature vector was then combined with EHR data to train a fully connected layer followed by two MTLR layers to estimate the risk. Finally, the risk predictions from the MTLR model and the CoxPH results are averaged to calculate the final risk scores.

\section{Results}
  We trained the models using all the training data before applying it on the HECKTOR test set. The concordance index (C-index), one of the metrics used to measure the performance of a prognosis model \cite{allen}, was used to report the results on the HECKTOR test dataset on each of the different models previously mentioned as shown in Table \ref{tab1}. The baseline model which only uses MTLR to estimate the risk has 0.66 C-index. Slight improvement on C-index was achieved when combining image features and EHR in the MTLR + Deep-CR model (C-index of 0.67). The results obtained using Deep Fusion (V1) has also achieved a C-index score of 0.67. However, the best estimation of risk was obtained using Deep Fusion (V2) with C-index of 0.72. 

\begin{table}

\caption{C-index scores obtained on the HECKTOR 2021 testing dataset.}\label{tab1}
\begin{tabularx}{\textwidth}{|X|p{3cm}|}
\hline
\textbf{Models} & \textbf{C-index}\\
\hline
MTLR (Baseline) & 0.66\\
MTLR + Deep-CR \cite{kim} &  0.67\\
MTLR + CoxPH + Deep Fusion (V1) & 0.67\\
MTLR + CoxPH + Deep Fusion (V2) & \textbf{0.72}\\
\hline
\end{tabularx}
\end{table}

\section{Discussion and Conclusion}
The results of the Deep Fusion (V1) suffered a low score of 0.67 C-index compared to the (V2). This augmentation-like approach of feeding CT, PET and Fused version into individual CNN architectures, combining the outputs and forwarding them into MTLR, then finally concatenating them with CoxPH results was hypothesized to yield better results. However, C-index of 0.67 is achieved compared to 0.72 in (V2). Multiple possibilities could have contributed to this discrepancy. First, the training of 3 different CNNs was not optimized to generate a well representative feature vector. This may have led to misleading feature vectors that make it hard to train a discriminative MTLR model. Second, the final aggregation of the output was merely concatenating the three feature vectors. Introducing more sophisticated aggregation of these feature vectors such as attention mechanism may improve the representation power in the latent space. 

In this paper, we introduced a machine learning model which uses EHR of patients along with a fused CT and PET images to perform prognosis for H\&N tumor patients. The model has achieved 0.72 C-index using an optimized network. This was part of the HECKTOR 2021 challenge which has only allowed a small number of submissions during the challenge period. This has limited this work on exploring other possible experiments and optimizations to the proposed models. Nevertheless, investigating different CNN architectures as well as ways to combine EHR and imaging data has the potential to improve the model.

%
%
%
\bibliographystyle{splncs04}
\bibliography{main}

\end{document}